\documentclass[twocolumn,showpacs,superscriptaddress, preprintnumbers , amsmath,amssymb,letterpaper]{revtex4-1}

\usepackage{graphicx}
\usepackage{dcolumn}
\usepackage{bm}
\usepackage{amsmath}
\usepackage{multirow}
\newcommand{\astra}{{\sc Astra }}

\usepackage[colorlinks=true]{hyperref}

\begin{document}
\preprint{FERMILAB -PUB-15-083-APC}
\date{\today}
\title{Tailored electron bunches with smooth current profiles for enhanced transformer ratios in beam-driven acceleration}

\author{F. Lemery} \affiliation{Northern Illinois Center for
Accelerator \& Detector Development and Department of Physics,
Northern Illinois University, DeKalb IL 60115,
USA} 
\author{P. Piot} \affiliation{Northern Illinois Center for
Accelerator \& Detector Development and Department of Physics,
Northern Illinois University, DeKalb IL 60115,
USA} \affiliation{Accelerator Physics Center, Fermi National
Accelerator Laboratory, Batavia, IL 60510, USA}

\begin{abstract}
Collinear high-gradient ${\cal O} (GV/m)$ beam-driven wakefield methods for charged-particle acceleration could be critical to the realization of  compact, cost-efficient,  
accelerators, e.g., in support of TeV-scale lepton colliders or multiple-user free-electron laser facilities. To make these options viable, the high 
accelerating fields need to be complemented with large transformer ratios $>2$, a parameter characterizing the efficiency of the energy transfer 
between a wakefield-exciting ``drive" bunch to an accelerated ``witness"  bunch. 
While several potential current distributions have been discussed, their practical realization appears challenging due to their often discontinuous nature. In this
paper we propose several alternative current profiles which are smooth which also lead to enhanced transformer ratios. We especially explore a laser-shaping method capable 
of  generating one the suggested distributions directly out of a photoinjector and discuss a linac concept that could possible drive a dielectric accelerator. 
\end{abstract}

\maketitle

%
%
\section{Introduction}
In beam-driven techniques, a high-charge ``drive'' bunch passes through a high-impedance medium and experiences a decelerating field~\cite{pisin,petra,gai}.
The resulting energy loss can be transferred to a properly delayed ``witness'' bunch trailing the drive bunch. A critical parameter associated 
to this class of acceleration method is the transformer ratio
\begin{eqnarray}
{\cal R}\equiv\left|\frac{E_+}{E_-}\right|,
\end{eqnarray}
where $E_+$ is the maximum accelerating field behind the drive bunch, and $E_-$ is the maximum decelerating field within the drive bunch. 

Generally the transformer ratio is limited to values ${\cal R} \leq 2$ due to the fundamental beam-loading theorem~\cite{wake}. However larger 
values can be produced using drive bunches with tailored (asymmetric) current profiles. Furthermore, it can be shown that both ${\cal R}$ and 
$E_+$ for a given charge are maximized when the decelerating field over the drive bunch is constant~\cite{tsakanov}. Additionally, bunch 
current profiles that minimize the accumulated energy spread within the drive bunch are desirable as they enable transport of the drive bunch 
over longer distances.\\

To date, several current profiles capable of generating transformer ratios ${\cal R} > 2$ have been proposed~\cite{tsakanov,powersTR,jing}. 
These include linearly ramped profiles combined with a  door-step or exponential initial distribution~\cite{bane}. 
More recently a piecewise ``double-triangle" current profile was suggested as an alternative with the advantage of being 
experimentally realizable~\cite{jiang}.  A main limitation common to all these shapes resides in their discontinuous character which make 
their experimental realization either challenging or relying on complicated beam-manipulation techniques~\cite{muggli,piotprstab11}. 
In addition these shapes are often foreseen to be formed in combination with an interceptive mask~\cite{yinesun,powerIPAC14} which add further challenges 
when combined with high-repetition-rate linacs~\cite{zholents14}. \\

In this paper we introduce several smooth current profiles which support large transformer ratios and lead to quasi-constant decelerating fields 
across the drive bunch.  We describe a simple scheme for realizing one of these shapes in a photoemission radiofrequecy (RF) electron source employing  
a shaped photocathode-laser pulse. Finally, we discuss a possible injector configuration that could form drive bunches consistent with the multi-user
free-electron laser (FEL) studied in Ref.~\cite{zholents14}.

\section{Smooth Shapes\label{sec:theory}}
For simplicity we consider a wakefield-assisting medium (e.g. a plasma or a dielectric-lined waveguide) that supports an axial wakefield described by the 
Green's function~\cite{wilson} 
\begin{equation}\label{eq:greens}
G(\zeta)=2\kappa \cos(k \zeta), 
\end{equation}
where $\kappa$ is the loss factor and $k \equiv2\pi /\lambda$ with $\lambda$ being the wavelength of the considered mode. Here $\zeta>0$ (in our convention) is the distance behind the source particle responsible for the wakefield. In this Section we do not specialize to any wakefield mechanism and recognize that, depending on the assisting medium  
used to excite the wakefield, many modes might be excited so that the Green's function would consequently consist of a summation over these modes.  

Given the Green's function, the voltage along and behind a bunch with axial charge distribution $S(z)$ can be obtained from the convolution~\cite{wilson}
\begin{equation}\label{eq:wakepot}
V(z)=\int_{-\infty}^z G(z-\zeta) S(\zeta) d\zeta.
\end{equation}
%
%

We take $S(z)$ to be non vanishing on two intervals $[0,\xi]$ and $[\xi, Z]$ and zero elsewhere. In our convention the bunch head starts at $z=0$ 
and the tail population lies at $z>0$. We also constrain our search to functions such that $S(z)$ and $S'(z)\equiv dS/dz$ are continuous at $z=\xi$. 
Introducing the function $f(z)$ (to be specified later), we write the charge distribution as 
\begin{equation}\label{eq:genramp}
S(z)= 
\begin{cases} f(z) & \text{if $0 \le z < \xi$,} 
\\
f'(\xi) z - f'(\xi) \xi + f(\xi) &\text{if $\xi\le z \le Z $,}
\\ 0 &\text{elsewhere.}
\end{cases}
\end{equation}
%
%
\subsection{Linear ramp with sinusoidal head \label{sec:sinramp}}
Based on our previous work~\cite{flemeryAAC14} we first consider the following function
\begin{eqnarray}\label{eq:sinramp}
f(z)=az+b \sin (qkz), 
\end{eqnarray}
where $a$ and $b$ are positive constants, $k$ is again the spatial frequency seen above, and $q>0$ is an integer. Consequently, using Eq.~\ref{eq:genramp}, the axial bunch profile is written as
\begin{equation}\label{eq:sinuramp}
S(z)= 
\begin{cases} az+b \sin (qkz )  & \text{if $0 \le z < \xi$,} 
\\
a x+b q k (x-\xi) \cos (q \xi k)\\
+b \sin (q \xi k) & \text{if $\xi\le z \le Z $,}
\\ 0 &\text{elsewhere.}
\end{cases}
\end{equation}
In this section we report only on solutions pertaining to $\xi = \lambda/2$.  Additional, albeit more complicated, solutions also exist for larger $\xi$; however, these 
solutions lead to additional oscillations which ultimately lowers the transformer ratio.\\

From Eq.~\ref{eq:wakepot}, the decelerating field then takes the form
\begin{widetext}
\begin{eqnarray}
E_-(z) &= \kappa
\begin{cases}
	\frac{\lambda}{\pi ^2}  \left(a \lambda  \sin ^2\left(\frac{\pi  z}{\lambda }\right)+\frac{\pi  b q \left(\cos \left(\frac{2 \pi  z}{\lambda }\right)-\cos \left(\frac{2 \pi  q z}{\lambda
   }\right)\right)}{q^2-1}\right) & z<\lambda/2\\
 \frac{\lambda  \left(\left(q^2-1\right) \left(a \lambda +2 \pi  b (-1)^q q\right)+\cos \left(\frac{2 \pi  z}{\lambda }\right) \left(2 \pi  b q \left((-1)^q q^2+1\right)-a
   \left(q^2-1\right) \lambda \right)\right)}{2 \pi ^2 \left(q^2-1\right)} & z \geq \lambda/2 \\
\end{cases} 
\end{eqnarray}
\end{widetext}

The oscillatory part in the tail ($\lambda/2 \leq z$) can be eliminated under the condition
\begin{equation}\label{eq:flatCond}
b = \frac{a \left(q^2-1\right) \lambda }{2 \pi  q \left((-1)^q q^2+1\right)},
\end{equation}

which leads to the following decelerating and accelerating fields respectively
\begin{eqnarray}
E_-(z) &=& \kappa
\begin{cases}
 \frac{a \lambda ^2 \left(2 (-1)^q q^2 \sin ^2\left(\frac{\pi  z}{\lambda
   }\right)-\cos \left(\frac{2 \pi  q z}{\lambda }\right)+1\right)}{2 \pi
   ^2 \left((-1)^q q^2+1\right)} & z<\lambda/2 \\
 \frac{a \left((-1)^q \left(2 q^2-1\right)+1\right) \lambda ^2}{2 \pi ^2
   \left((-1)^q q^2+1\right)} & z\geq \lambda/2 \\
\end{cases}
\end{eqnarray}	
\begin{widetext}
\begin{eqnarray}
	\begin{split}	
	E_+(z)& = \int_0^{Z\equiv N\lambda} s(z')\omega(z-z') dz' \\
	& =\kappa \frac{a \lambda ^2 \left(\pi  \left((-1)^q \left((4 N-1) q^2-2
   N+1\right)+2 N\right) \sin \left(2 \pi  \left(N-\frac{z}{\lambda
   }\right)\right)+\left((-1)^q \left(2 q^2-1\right)+1\right) \cos \left(2
   \pi  \left(N-\frac{z}{\lambda }\right)\right)\right)}{2 \pi ^2
   \left((-1)^q q^2+1\right)}.
 	\end{split}
\end{eqnarray}
\end{widetext}
Finally, the transformer ratio can be calculated by taking the ratio of the maximum accelerating field (see Appendix~\ref{app:A}) over the maximum decelerating field which yields

\begin{widetext}
\begin{eqnarray}
	{\cal R} = \frac{\sqrt{\pi ^2 \left((-1)^q \left((4 N-1) q^2-2 N+1\right)+2 N\right)^2+\left((-1)^q \left(2
   q^2-1\right)+1\right)^2}}{(-1)^q \left(2 q^2-1\right)+1}.	
\end{eqnarray}
\end{widetext}

\begin{figure}[ttt!!!!!!!]
\centering
\includegraphics[width=0.48\textwidth]{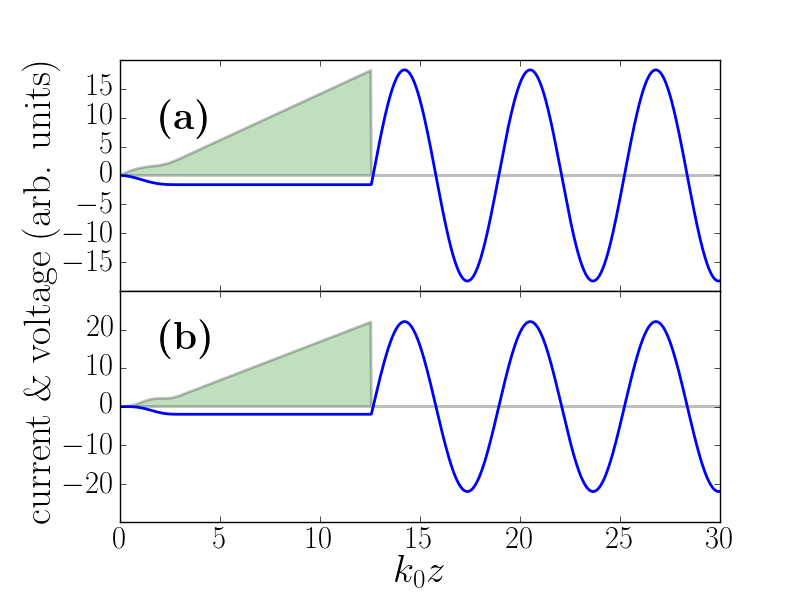}
\caption{Example of current profiles described by Eq.~\ref{eq:sinuramp} (shaded line) with the corresponding induced voltages. The parameters are $n=0$, $N=5,$ and plots (a) and (b) respectively correspond to the cases $q=2$  and  $q=3$. The head of the bunch is at $k_0 z=0$. \label{fig:sinRamp}}
\end{figure}

Two sets of solutions occur for even and odd $q$ which can be interpreted as a phase shift in the oscillatory part.  Additionally, larger multiples of even and odd $q$ 
lead to more oscillations in the head which ultimately reduce the transformer ratio.  In Fig.~\ref{fig:sinRamp} we illustrate the simplest even (a) and odd (b) 
solutions corresponding to $q=2$ and $q=3$ respectively.

\subsection{Linear ramp with parabolic head\label{sec:theory-quad}}
We now consider an even simpler quadratic shape which was inspired by our previous work~\cite{flemeryAAC14,andonian} 
  
\begin{eqnarray}\label{eq:sinramp}
f(z)=az^2, 
\end{eqnarray}
which leads to the current profile 
\begin{equation}\label{eq:quadratic}
S(z)= 
\begin{cases} az^2  & \text{if $0 \le z < \xi$,} 
\\
2a\xi z - a \xi^2 &\text{if $\xi\le z \le Z $,}
\\ 0 &\text{elsewhere.}
\end{cases}
\end{equation}
The resulting decelerating field within the bunch is 
\begin{eqnarray}
E_-(z)= 2 \kappa 
\begin{cases} 
-2a\frac{\sin(k z)-k z}{k^3}  & \text{if $0 \le z < \xi$,}  \\
2a\frac{\sin[k(z-\xi)]-\sin(kz)+2k \xi}{k^3} &\text{if $\xi\le z \le Z $,}
\\ 0 &\text{elsewhere.} 
\end{cases} \nonumber
\end{eqnarray}
Again, the decelerating field can be made constant for $z \in [\xi,Z]$ when $\xi=  \nu \lambda$ with $\nu \in \mathbb{N}$. In such a case the previous equation simplifies to  
\begin{eqnarray}
E_-(z)= 2\kappa
\begin{cases} 
-2a\frac{\sin(k z)-k z}{k^3}  & \text{if $0 \le z < \nu \lambda$,}  \\
  \frac{4 \pi a \nu}{k^3 } &\text{if $\nu\lambda \le z \le Z $,}
\\ 0 &\text{elsewhere.}
\end{cases}\nonumber
\end{eqnarray}
\begin{figure}[hhhhhhhh!!!!!!!]
\centering
\includegraphics[width=0.45\textwidth]{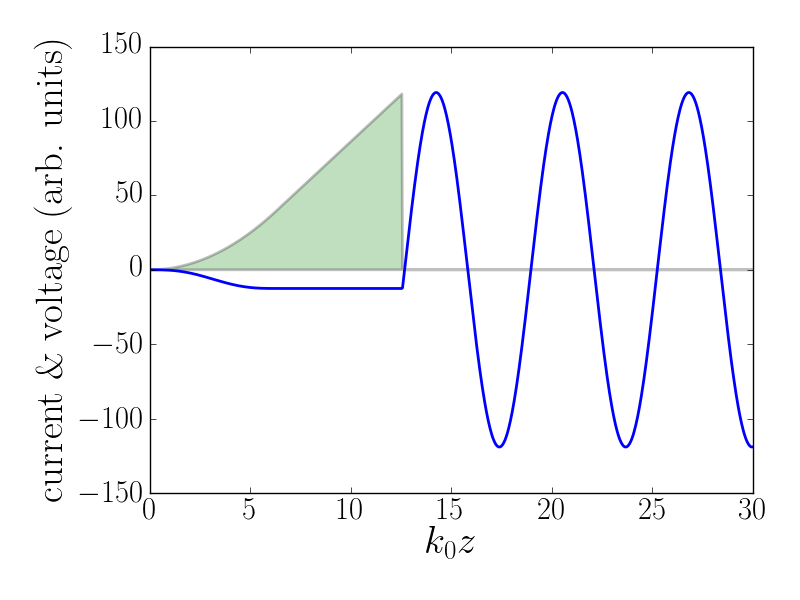}
\caption{Example of ideal ``quadratic" current profiles given by Eq.~\ref{eq:quadratic} (shaded line) with corresponding induced voltage. The parameters are $\nu=1$, and 
$N=5$. The head of the bunch is at $k_0 z=0$. \label{fig:quadratic}}
\end{figure}
%
The accelerating field trailing the bunch $(z> Z\equiv N\lambda_0)$ is 
\begin{eqnarray}
E_+(z)&=& - \frac{8 \pi \nu a \kappa }{k^3} [ \pi (2N -\nu) \sin (k z)  \\ \nonumber 
 && - \cos (k z) ], 
\end{eqnarray}
yielding the transformer ratio 
\begin{eqnarray}
{\cal R} &=&  [1+\pi^2(2N-\nu)^2]^{1/2}. 
\end{eqnarray}
In Fig.~\ref{fig:quadratic} we illustrate an example of the quadratic shape (green trace) as well as its corresponding longitudinal electric field (blue trace)
for $\nu=1$ and $N=5$.

%

%
%
%
\subsection{Comparison with other shapes\label{sec:theory-comp}}
\begin{figure}[hhhhhhhh!!!!!!!]
\centering
\includegraphics[width=0.50\textwidth]{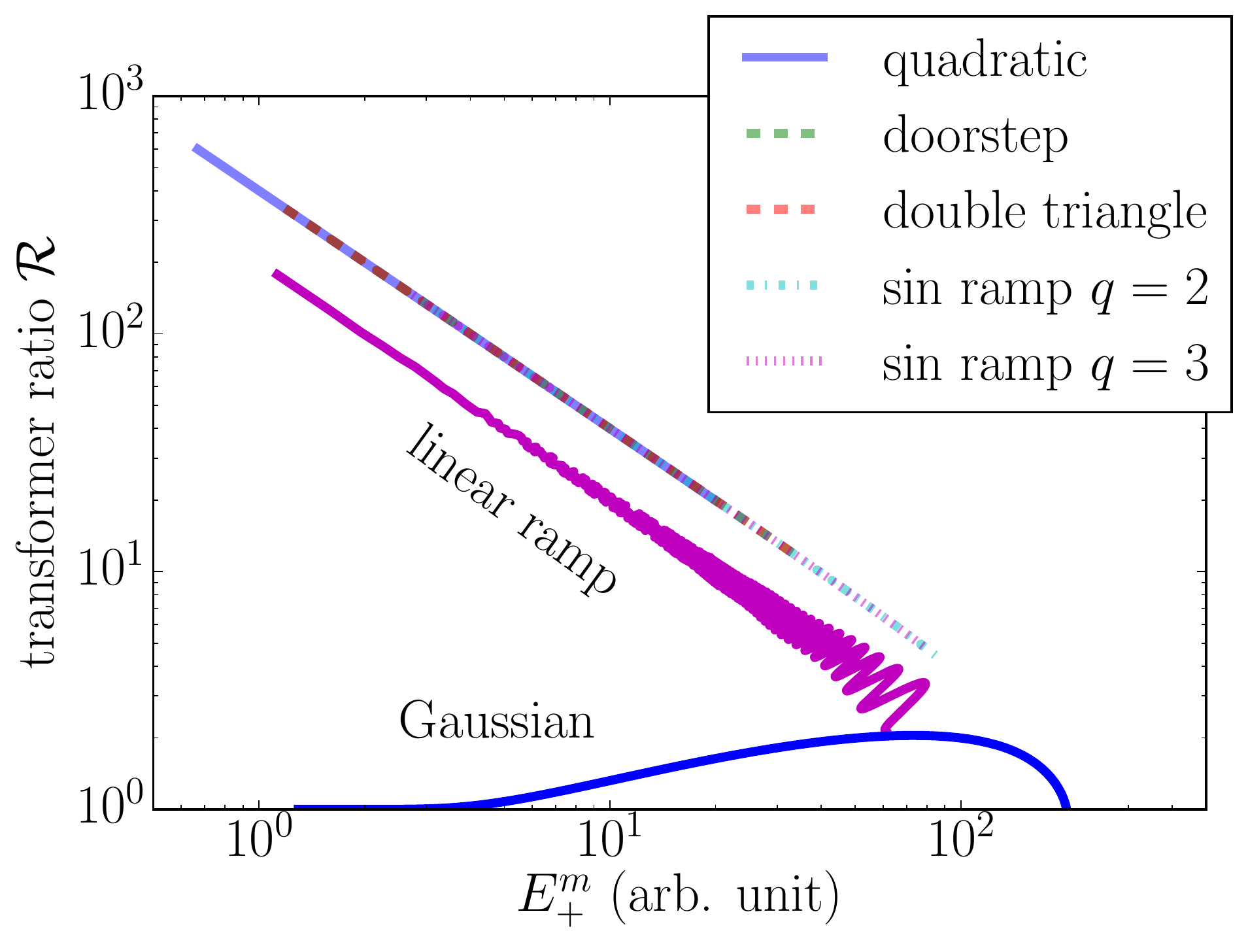}
\caption{Tradeoff curves between ${\cal R}$ and $E_+^m$ for the current profiles listed in Tab.~\ref{tab:Rcompare}. The ``quadratic" and ``sin ramps" respectively 
correspond to the distributions proposed in Sections~\ref{sec:theory-quad} and~\ref{sec:sinramp}. The Gaussian and ramp distributions are displayed for 
comparison. \label{fig:EvRanalytic}}
\end{figure}

We now turn to compare the smooth longitudinal shapes from the previous Section with the doorstep~\cite{bane}  and double-triangle~\cite{jiang} which also
provide constant decelerating fields over the bunch-tail (see Appendix~\ref{app:B} for our formulation of these distributions).
For a fair comparison, we stress the importance of comparing the various current profiles with equal charge.  
Consequently, we normalize each of the current profile to the same bunch charge  
\begin{equation}\label{eq:normalization}
	Q = \int_0^{Z\equiv N\lambda} dz S(z,a);
\end{equation}
where $a$ is the scaling parameter associated with each bunch shape (see Section~\ref{sec:theory} and Appendix~\ref{app:B}), 
and $N\lambda$ is the total bunch length which is assumed to be larger than the given shape's bunch-head length ($N\lambda>\xi$). 
For each distribution, the charge normalization generates a relationship between $a$ and $N \lambda$ which enables us to rexpress  
${\cal R}$ in terms of $Q$ and $a$. In Tab.~\ref{tab:Rcompare} we tabulate the analytical 
results for ${\cal R}(N)$ (the conventional notation ~\cite{bane,jiang}) and ${\cal R}(Q,a)$, and also list the maximum decelerating field $E_-^m$ for each distribution.
Additionally in Fig.~\ref{fig:EvRanalytic} we illustrate these results in a log-log plot where, for each distribution, the scaling parameter ($a$) 
was varied for a fixed charge and wavelength. To complete our comparison we also added the linear-ramp and Gaussian distributions.

The results indicate that all of the distributions with constant decelerating fields over the bunch-tail `live' on the same curve;
additionally, by varying the scaling parameter $a$ for a distribution, you can shift a distribution to have a larger (resp. smaller) 
${\cal R}$ (resp. $E_+$) and vice-versa.  Ultimately, this suggests that the distribution which is simplest to make is as useful as any other and
it can be scaled accordingly (${\cal R}$, $E_+$) for a specific application. These results confirm our previous studies regarding the numerical
investigation of the trade-off between ${\cal R}$ and $E_+$~\cite{lemeryshape}.

\begin{table*}[t]
    \caption{\label{tab:Rcompare}
    Table comparing several different proposed drive bunch distributions as a function of bunch length and charge.  Additionally,
	the maximum decelerating field ($E_-^m$) is shown for each distribution.  Here we consider $\kappa =1$.}
    \begin{ruledtabular}
        \begin{tabular}{lcc r}
            distribution & ${\cal R}(N)$  & ${\cal R}(Q)$  & $E_-^m$  \\
            \hline
            doorstep~\cite{bane}  & $\sqrt{1+(1-\pi/2 + 2\pi N)^2}$ & $\sqrt{2+\pi(\frac{4Q}{a\lambda}-1)}$ & $\frac{a\lambda}{\pi}$ \\
            double triangle~\cite{jiang}  & $\sqrt{1+(2\pi N -1)^2}$ & $\sqrt{2+\pi(\frac{4Q}{a\lambda}-1)}$ & $\frac{a\lambda}{\pi}$\\
			sin ($q=2$)  & $\frac{1}{8} \sqrt{\pi ^2 (3-16 N)^2+64}$ & $\frac{1}{8}\sqrt{64-15\pi^2+\frac{48\pi Q}{a\lambda}}$ & $\frac{16a\lambda}{3\pi}$\\
			sin ($q=3$)  & $\frac{1}{2} \sqrt{\pi ^2 (1-4 N)^2+4}$ & $\frac{1}{6}\sqrt{44-9\pi^2+\frac{24\pi Q}{a\lambda}}$ & $\frac{6a\lambda}{\pi}$\\
            quadratic  & $\sqrt{1 + \pi^2 (2N-1)^2}$ & $\sqrt{1+\pi^2(\frac{4Q}{a\lambda^3}-\frac{1}{3})}$ & $\frac{a \lambda ^3}{\pi ^2}$
        \end{tabular}
    \end{ruledtabular}
\end{table*}

\section{Photoemission of optimal shapes via laser-shaping} 
\begin{figure}[hhhhhhhh!!!!!!!]
\centering
\includegraphics[width=0.45\textwidth]{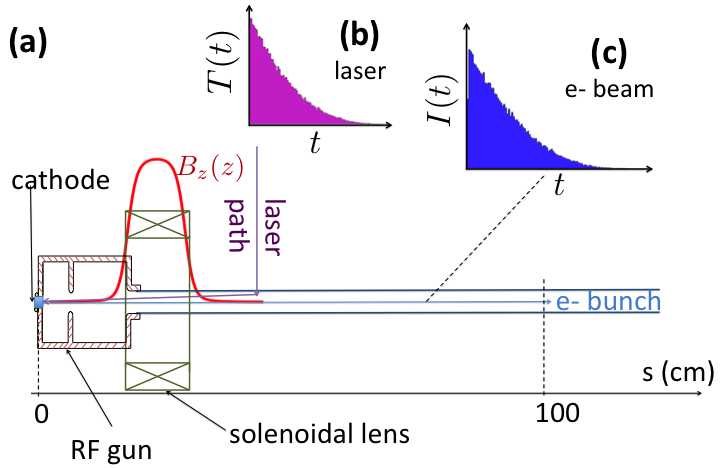}
\caption{Configuration used for the pulse-shaping simulations using a S-band RF gun (a). A temporally shaped laser pulse (b) is optimized to result 
in a photo-emitted electron-beam with current profile (c) having features similar to the distribution discussed in Sec.~\ref{sec:theory-quad}.  In insets 
(a) and (b) the tail of the bunch is at $t=0$.
\label{fig:setup}}
\end{figure}

In this section we investigate the realization of the quadratic distribution discussed in Section~\ref{sec:theory} by longitudinally tailoring
a laser pulse impinging on a photocathode in a photoinjector.  The resulting electron distribution is then accelerated in an RF-gun and expands via space charge 
forces.  If the charge density of the emanating electron bunch is sufficiently low, the resulting distribution will be relativistically preserved 
through a drift; however for larger charge densities, the original longitudinal distribution will morph according to the integrated space charge 
forces inside the bunch.  
The setup we consider throughout this section is depicted in Fig.~\ref{fig:setup} and consists of a typical $1+\frac{1}{2}$-cell BNL/SLAC/UCLA 
S-band RF-gun operating at 2.856~GHz surrounded by a solenoidal lens~\cite{slacgun}. The large ($\sim 140$~MV/m) acceleration gradients in the gun
help preserve larger charge densities compared with e.g. L-band guns.  The simulations are carried with 
\astra~\cite{astra}, a particle-in-cell beam-dynamics program that includes a quasi-static cylindrically-symmetric space charge algorithm. 
The simulation also includes the image-charge effect which arises during the photoemission process, in our simulations the electron bunch 
is represented by 200,000 macro-particles. 
\subsection{Case of an ideal laser-shaping technique}

We base our approach on Ref.~\cite{flIPAC14} where we developed a simple 1D space charge model to investigate the expansion forces in various distributions;
for polynomial ($x^{\sim2}$) distributions, we observed that the relatively small fields in the bunch-head will essentially preserve the local longitudinal form.
In the rear portion of the bunch however, there is an asymmetrical blowout which has proper sign to possibly lead
to a linear-like tail which is required for the quadratic-ramp.

We consider a laser intensity distribution of the form $I(r,t)=T(t)R(r)$, where $T(t)$ is the temporal profile and $R(r)$ the transverse laser envelope. 
In our previous studies we used a radially uniform transverse distribution and explored polynomial and exponential forms for $T(t)$. In this section we 
report on the performance of the polynomial distribution given by
\begin{equation}\label{eq:exp}
	T(t) = T_0 t^{\alpha} H(\tau-t),
\end{equation}
where $T_0$ is a normalization constant, $\alpha>0$ is the polynomial power, $\tau$ is the ending time of the pulse, and $H(t)$ is the Heaviside function.  

The exponent $\alpha$ greatly influences the space-charge fields. Large values of $\alpha$ (e.g $\alpha\gtrsim 5$) lead to large space-charge forces
which results in a uniformly filled ellipsoidal distribution~\cite{luiten}.  Alternatively, smaller values of $\alpha$ ($\alpha \lesssim3$) lead to a
more uniform evolution of the bunch dynamics due to the increased uniformity of the field over the bunch.
Additionally,  the transverse spot size of the laser pulse on the photocathode also controls the longitudinal electric fields but also influences the  
transverse ``thermal" emittances.  It is also possible to reduce the electric fields and the associated blowout rate by using longer laser pulses; in
this scenario, the resulting electron bunch will evolve at a slower rate but the resulting bunch distribution will have a smaller peak current compared to 
when starting with smaller values of $\tau$. A smaller current will  impact the performances of the wakefield accelerator (or require the implementation of a 
longitudinal compression scheme).  Finally, it would also be possible to use a longer, e.g.  $2+\frac{1}{2}$-cell, RF gun or another acceleration cavity in close 
proximity to the gun to preserve larger charge densities which could effectively alleviate the need for a bunch compressor to drive large accelerating fields in 
the subsequent wakefield accelerator.

\begin{figure}[hhhhhhhh!!!!!!!]
\centering
\includegraphics[width=0.45\textwidth]{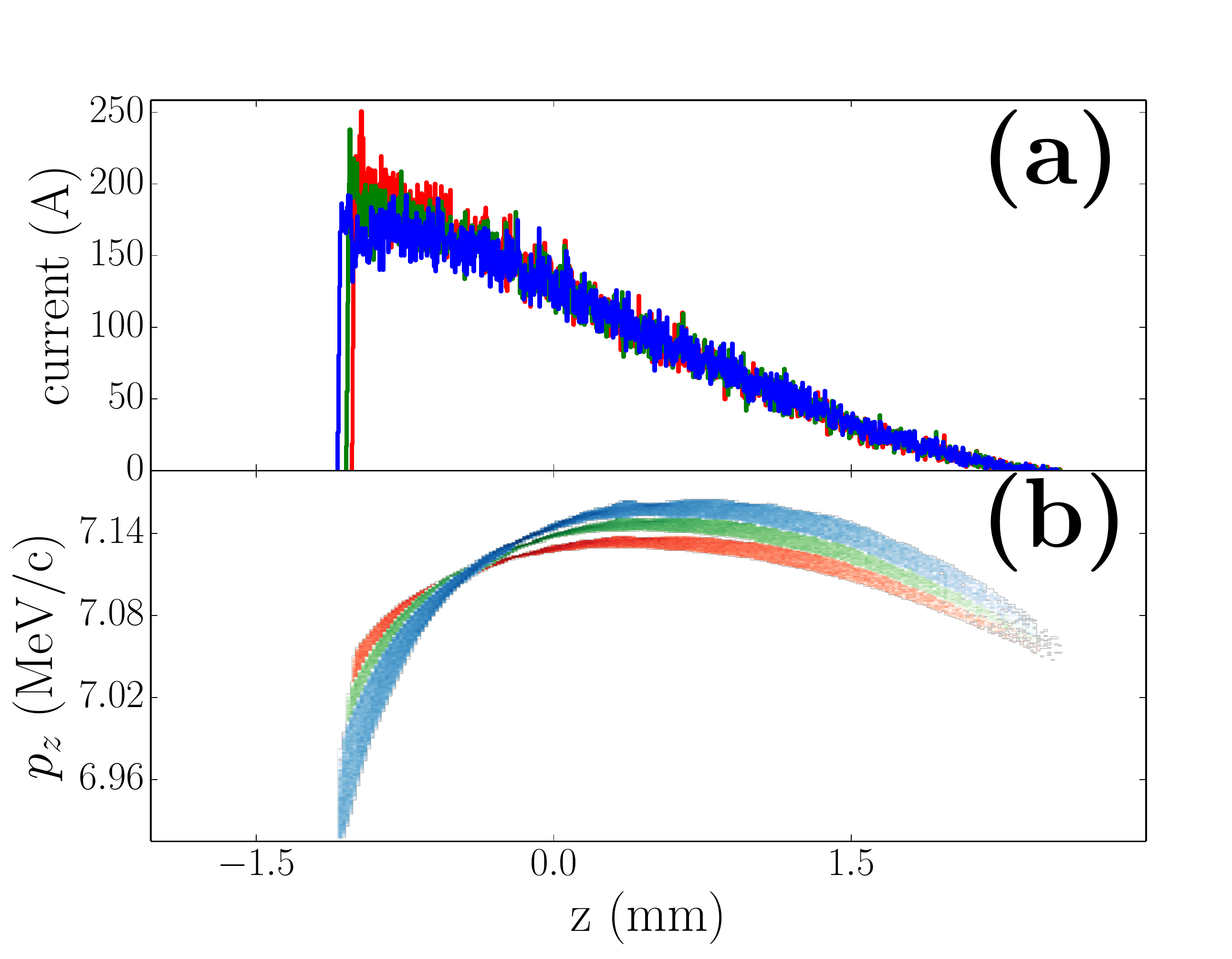}
\includegraphics[width=0.45\textwidth]{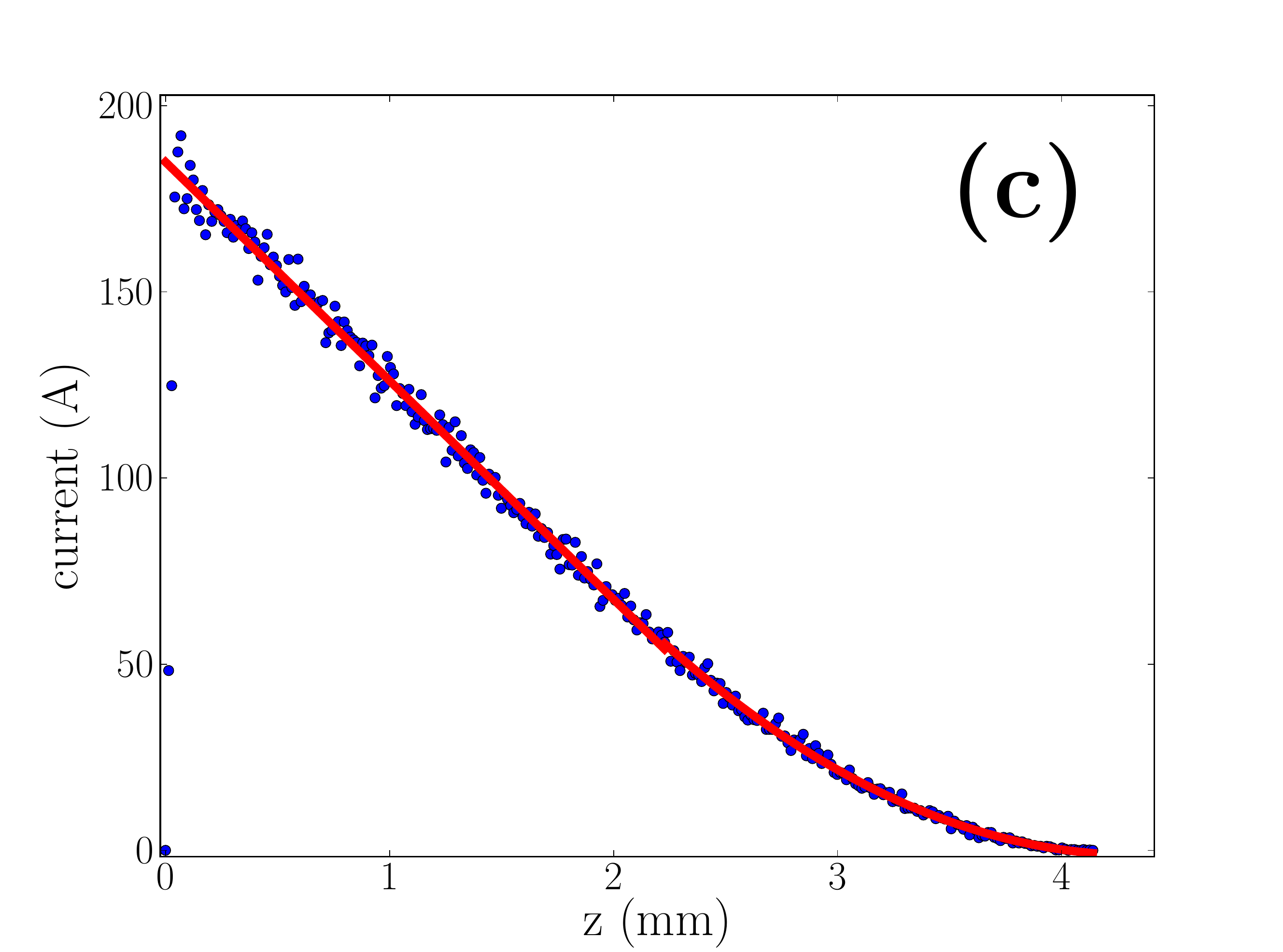}
\caption{Evolution of the electron-bunch current (a) and longitudinal phase space (b)  along the beamline at 20 (red), 60 (green), and 100~cm (blue) from the 
photocathode surface and (c) comparison of the current profile numerically simulated at $s=50$~cm (red trace) with a fit to equation Eq.~\ref{eq:quadratic} (blue 
lines). The head of the bunch is at large values of $z$.
\label{fig:polyEvos}}
\end{figure}

Figure~\ref{fig:polyEvos}  shows simulated longitudinal phase space snapshots and corresponding currents at different axial locations downstream of the gun for 
a 1-nC bunch. For this simulation a 1-mm rms laser spot size on the photocathode was used. The initial laser distribution was described by Eq.~\ref{eq:exp} with 
$\alpha=2$ and $\tau=15$~ps. A fit of the current distribution at $s=50$~cm from the photocathode is shown in Fig~\ref{fig:polyEvos} and indicates 
that the final electron bunch distribution is indeed accurately described by Eq.~\ref{eq:quadratic}. 


\subsection{Limitation of a practical laser shaping technique} 
As a first step toward a realistic model for the achievable shaped we consider the photoemission process to be resulting from frequency tripling 
of a $\lambda_0=800$-nm amplifier infrared (IR) pulse impinging a fast-response time cathode (with typical work functions corresponding to ultraviolet 
photon energy $\sim \lambda_0/3$). Such a setup is commonly used in RF photoinjectors such as the one 
discussed in the previous sections.  We further assume that the frequency up-conversion 
process does not affect the original laser's temporal shape (e.g. the UV-pulse temporal shape is identical to the IR-pulse temporal shape). Under such 
an assumption, the formation of the ideal temporal shape discussed in the previous Section is limited by the finite laser bandwidth and frequency 
response of the shaping process. \\

We consider an incoming amplified IR pulse with intensity $I_{in} (r,t)=I_0(r) \mbox{sech}^2(t/\tau)$ downstream of the last-stage amplification, where 
$\tau$ is the laser pulse duration. We model the IR pulse laser-shaping process via the convolution $I_{out}(r,t)=\int_{-\infty}^{+\infty} I_{in} (r,t-t') R(t') dt'$ 
where $I_{out}(t)$ and $R(t)$ represent the shaped-pulse intensity and response function of the shaping method respectively. 

Given the desired output shape and incoming laser pulse profile, the response function of the shaping process has to be set to satisfy~\cite{wiener}
\begin{eqnarray}
\widetilde{R}(\omega) = \frac{\widetilde{I}_{out}(\omega)}{\widetilde{I}_{in}(\omega)}, 
\end{eqnarray}
where the upper tilde represents the Fourier transformation $\widetilde{f}(\omega)=\int_{-\infty}^{+\infty} f(t) e^{i\omega t}$.  In practice $I_{in}(\omega)$ is 
defined over a finite range of frequency $\omega = \omega_0 \pm \frac{\delta\omega}{2}$ where $\omega_0\equiv \frac{2\pi c}{\lambda_0}$ is the central laser frequency and $\delta \omega\equiv\frac{\omega_0}{\lambda_0}\delta \lambda$ is  the laser pulse bandwidth ($\delta \lambda$ is the wavelength span of the pulse spectrum). 

\begin{figure}[hhhhhhhh!!!!!!!]
\centering
\includegraphics[width=0.48\textwidth]{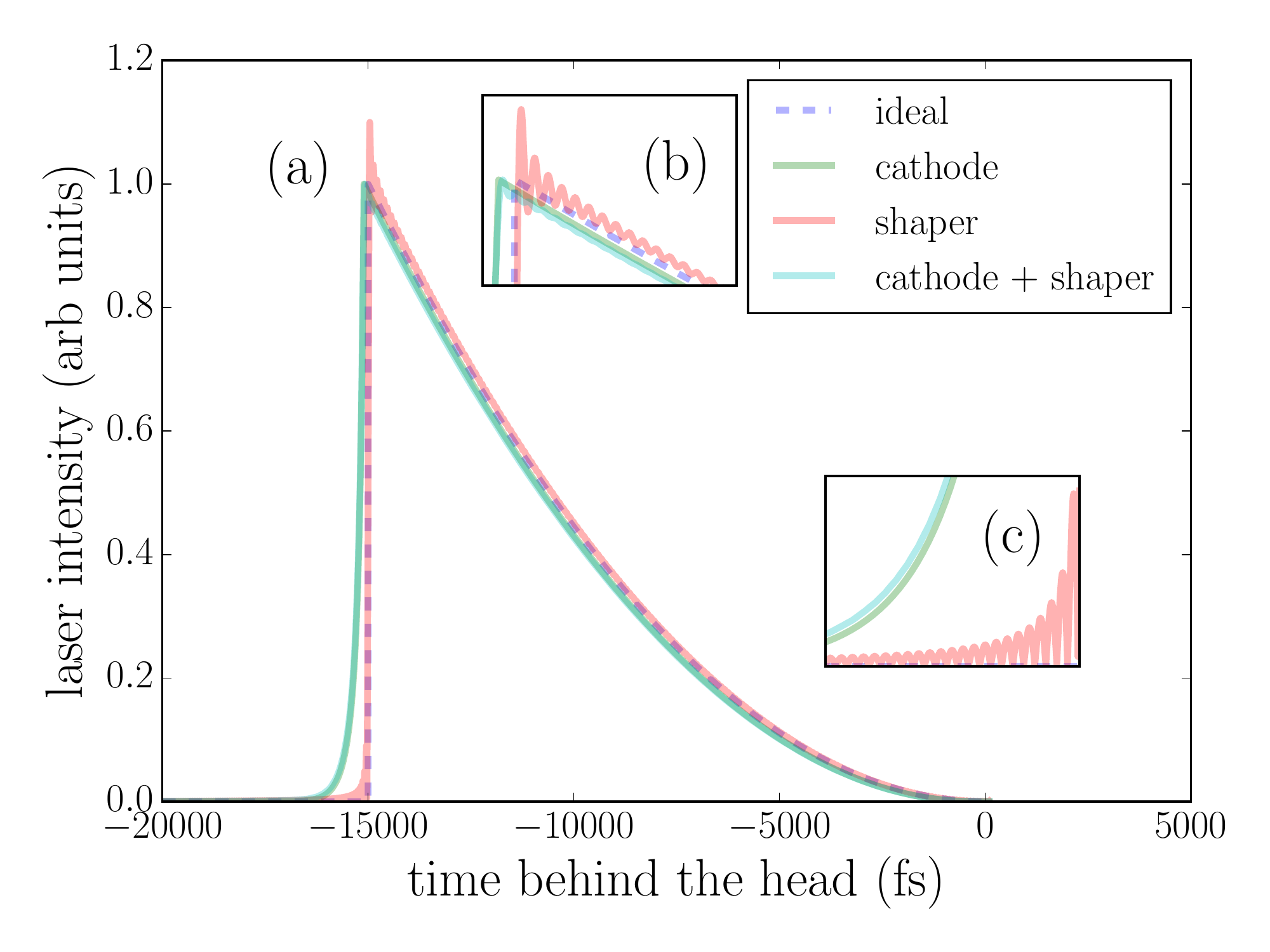}
\caption{Comparison of nominal (``ideal") shape with the shapes achieved when taking into account the photoemission response time (``cathode"), 
	the laser-pulse-shaping  finite bandwidth (``shaping") and both effects (``cathode + shaper"). The ideal laser temporal profile is described by Eq.~\ref{eq:exp} with $\alpha=2$ and 
$\tau=15$~ps. Insets (b) and (c) are zooms of the areas $t\in[-15200, -13600]$~ fs (peak location) and  $t\in[-16000, -15020]$~ fs (left edge of the profile) respectively. 
The head of the laser pulse is at $t=0$. 
\label{fig:shapelim}}
\end{figure}

\begin{figure}[hhhhhhhh!!!!!!!]
\centering
\includegraphics[width=0.48\textwidth]{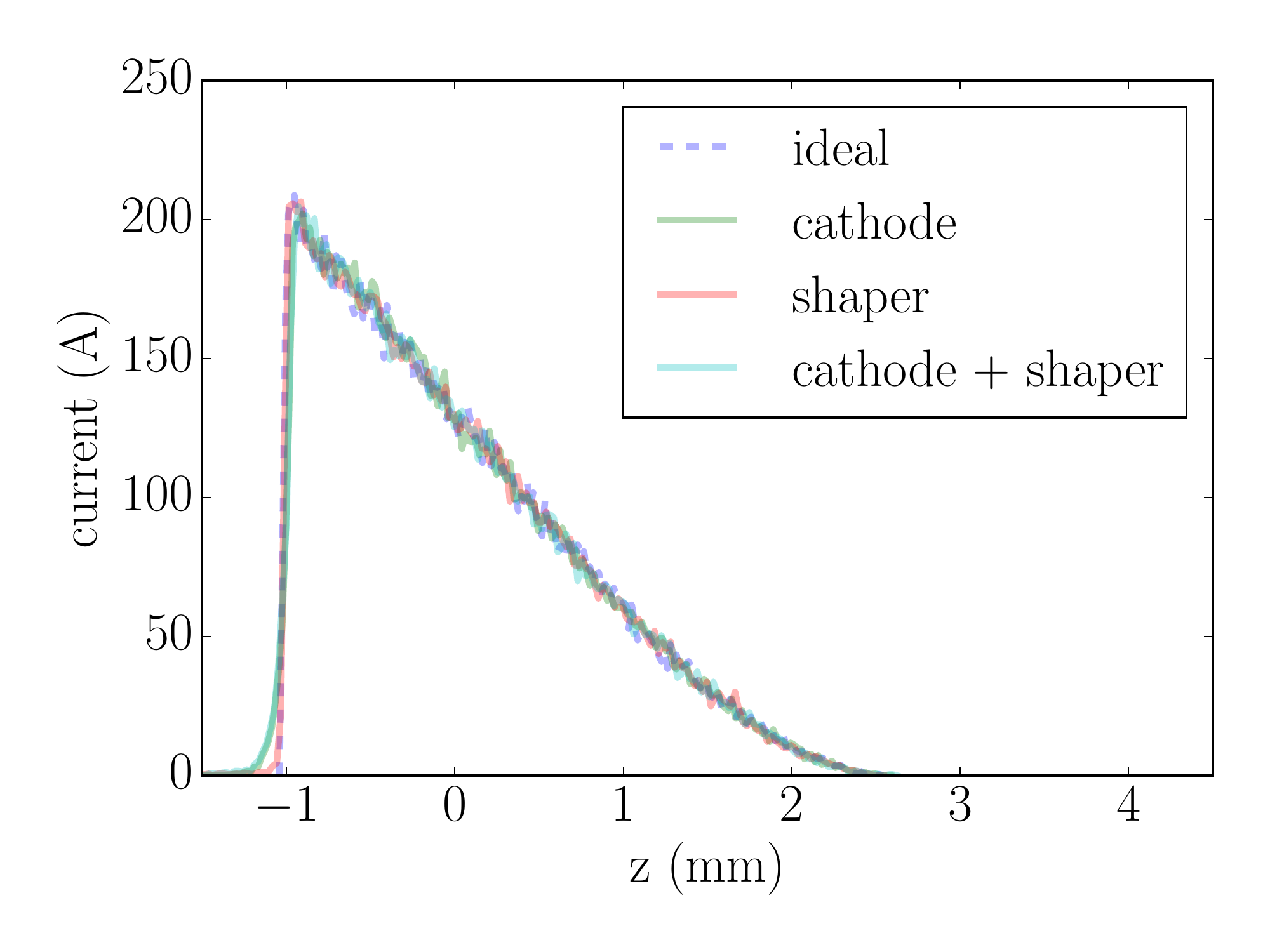}
\caption{Comparison of the final electron-bunch current at $s=50$~cm from the cathode surface for the four cases considered in Fig.~\ref{fig:shapelim}. The ``cathode" and ``shaper" 
respectively correspond to the inclusion of the cathode response time and shaper bandwidth limitation in the initial particle distribution at $s=0$ while the ideal case is given by 
by Eq.~\ref{eq:exp} with $\alpha=2$ and $\tau=15$~ps. The head of the bunch corresponds to $z>0$. 
\label{fig:shapefinal}}
\end{figure}

The typical shape considered in the previous section after laser shaping is shown in Fig.~\ref{fig:shapelim}; the limited 
bandwidth has very little effect except for the well-known ringing effect at the sharp discontinuities~\cite{gibbs}; see Fig.~\ref{fig:shapelim} (b) and (c). \\
Another potential limitation to our shaping scheme arises with a high-efficiency (semiconductor) photocathode. We consider as an example the case of Cs$_2$Te 
photocathodes because of their wide use in high-current photoinjectors.  The response-time limitation is investigated using the parameterized impulsional time
response of Cs$_2$Te described in Ref.~\cite{piotellipsoidal} based on numerical simulations presented in Ref.~\cite{ferrini}. The impulsional response is convolved 
with the distribution used in the  previous section and the results are gathered in Fig.~\ref{fig:shapelim}. Again this effect appears to be marginal.  For the sake of 
completeness, the various profiles shown in Fig.~\ref{fig:shapelim} are tracked with {\sc astra} and the final current distributions at $s=50$~cm are found to be 
indiscernibly close to the ideal shape considered in the previous Section; see Fig.~\ref{fig:shapefinal}. Such a result gives further confidence in the proposed shaping 
approach. 
%

%
\section{Formation of  high-energy  tailored bunches for a DWFA LINAc\label{sec:linac}} 
\begin{figure}[hhhhhhhh!!!!!!!]
\centering
\includegraphics[width=0.48\textwidth]{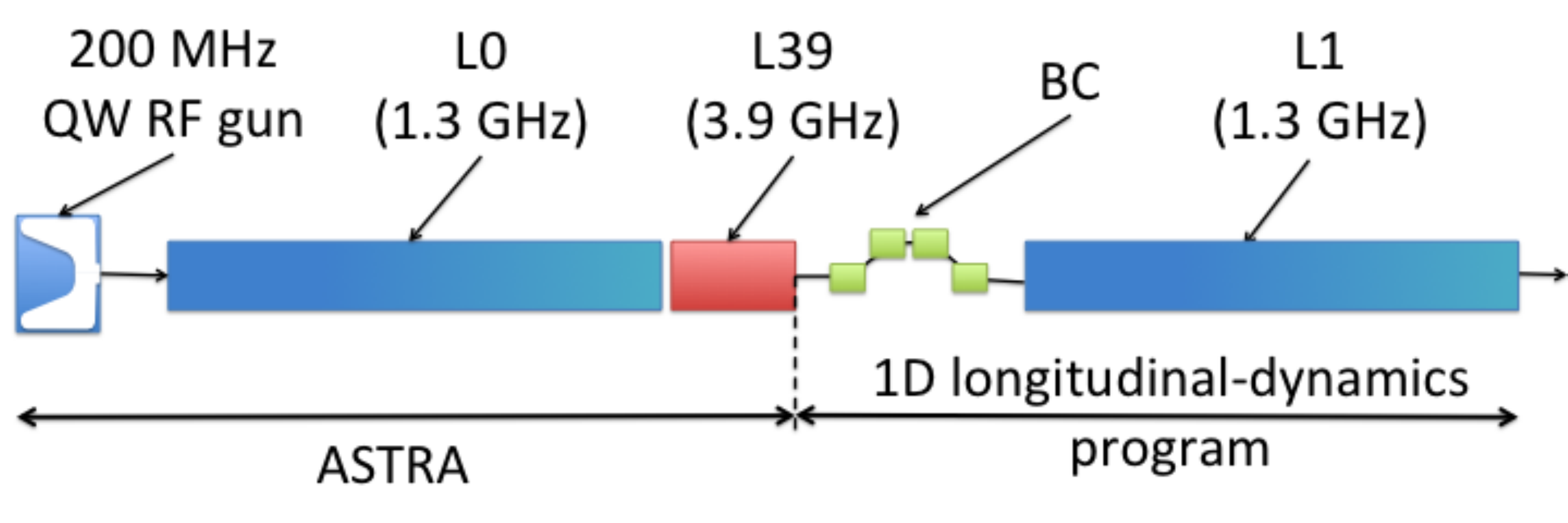}
\caption{Block diagram of the accelerator configuration explored for the formation of high-energy ramped bunches. The legend is as follows: ``QW" stands for quarter-wave, ``L0" and ``L1" are
standard 1.3-GHz cryomodule equipped with 8 TESLA-type SCRF cavities, ``L39" is a cryomodule consisting of four 3.9-GHz cavities, and "BC" is a magnetic bunch compressor.
\label{fig:DWFAdriver}}
\end{figure}

We finally investigate the combination of the tailored current-profile generation scheme with subsequent acceleration in a linac located downstream 
of the RF gun. Such a configuration could be useful to form tailored relativistic electron bunches for direct injection in wakefield-acceleration structures.  For this 
example, we consider a high-repetition drive bunch with parameters consistent with a recently proposed beam-driven accelerator for a short-wavelength free-electron
laser (FEL)~\cite{zholents14}.  We adopt a different approach than Ref~\cite{zholents14} and instead choose a 1.3-GHz superconducting RF (SCRF) linac (L0 and L1) composed of TESLA 
cavities~\cite{aunes} coupled to a quarter-wave 200-MHz SCRF gun~\cite{bobgun,JoeResults} originally designed for the WiFEL project~\cite{WiFEL}; see diagram in Fig.~\ref{fig:DWFAdriver}. 
The accelerator also includes a 3.9-GHz accelerating cavity (L39) section to remove nonlinearities in the longitudinal phase space~\cite{piotTESLANote,39GHz}. For this study we 
explored the use of polynomial laser profile described by Eq.~\ref{eq:exp} and let $\alpha$ and $\tau$ as free parameters.\\

\begin{figure}[hhhhhhhh!!!!!!!]
\centering
\includegraphics[width=0.48\textwidth]{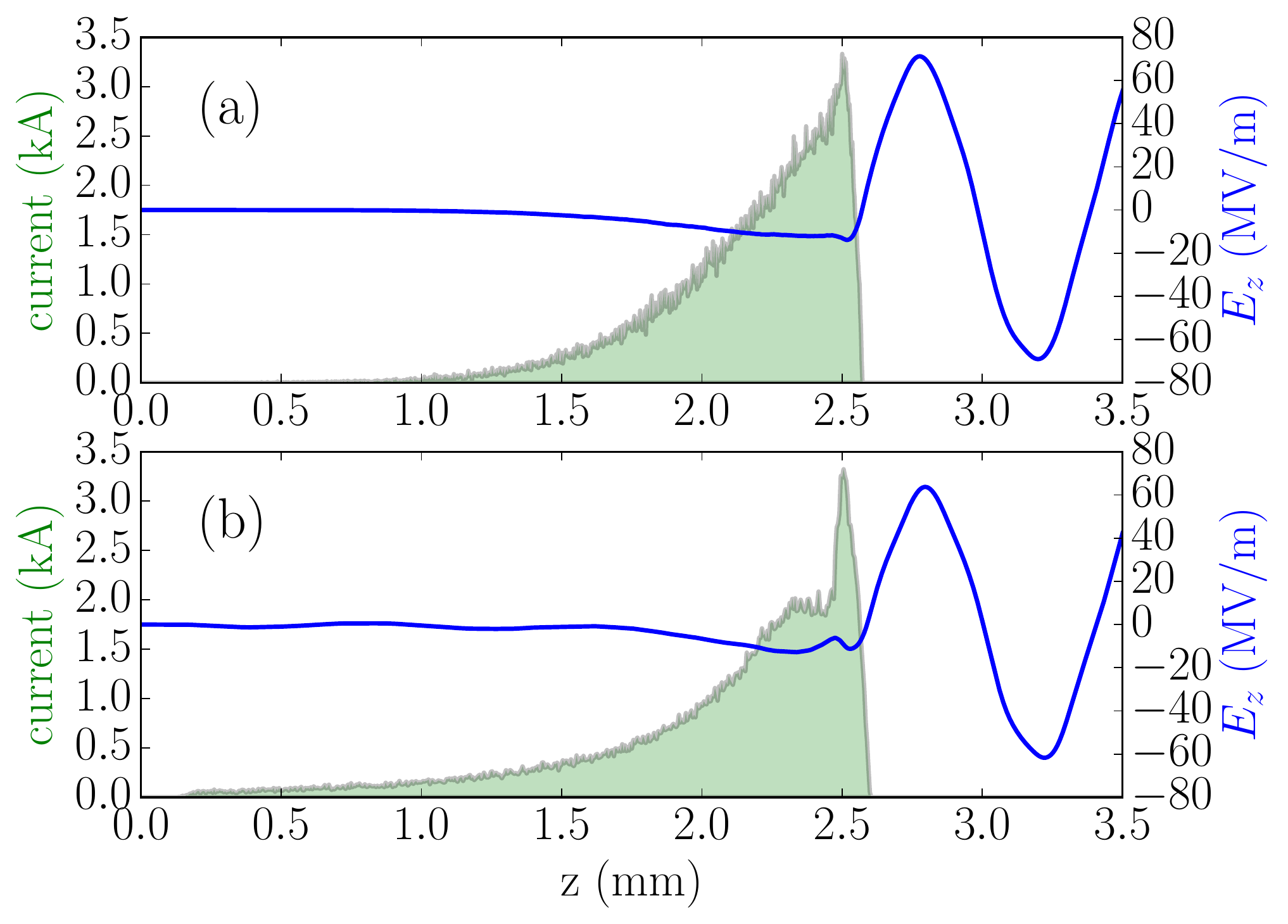}
\caption{Final current distribution (green shaded area) and associated wakefield (blue traces) for the ``ideal" (a) and ``realistic" (b) cases
of compression discussed in the text. The head of the bunch corresponds to $z=0$ \label{fig:finalwake}}.
\end{figure}

The laser-profile parameters and accelerator settings were optimized using a genetic optimizer~\cite{geneticoptimizer} to result in a final 
distribution with current profile consistent to achieve a high transformer ratio. The optimized accelerator settings are summarized in 
Tab.~\ref{tab:dwfaFEL}. In our optimization, we chose the wakefield structure to be a dielectric-lined 
waveguide with parameters tabulated in Tab.~\ref{tab:dlwperf} and we introduce a longitudinal scaling factor $\eta$ as free parameter such that 
 the axial coordinate is scaled following $z \rightarrow z'= \eta z$. The optimization converged to a value $\eta =0.16$.  The obtained wakefield 
 and scaled shape are shown in Fig.~\ref{fig:finalwake} (a). For the wakefield calculations we followed the formalism detailed in Ref.~\cite{rosinggai}  and use the 
 first  four modes in the wakepotential used for the beam dynamics simulations.   

\begin{table}[tt!]
\caption{\label{tab:dwfaFEL} Optimized settings for the accelerator parameters needed to produce and accelerate a ramp bunch to $\sim 200$~MeV.  
The parameter $\alpha$ and $\tau$ are defined in Eq.~\ref{eq:exp}. }
\begin{center}
\begin{tabular}{l c c  }
\hline \hline  parameter         &   value       & units  \\ \hline
laser rms spot size  $\sigma_r$    &    2.5 &  mm    \\
laser ramp $\alpha$ parameter & 19.86 & $-$ \\
laser ramp duration $\tau$ & 96.8 & ps \\
bunch charge $Q$ & 5 & nC \\ 
peak E-field on cathode   &    40  &  MV/m    \\
laser injection phase     &    71.0 &  deg (200 MHz)    \\
gun output beam momentum & 5.15 & MeV/c \\
acc. voltage L0   & 165 & MV/m  \\
off-crest phase L0  & -12.35 & deg (1.3 GHz) \\
acc. voltage L39  & 24.1 & MV  \\
off-crest phase L39  & -192.35 & deg (3.9 GHz) \\
beam momentum after L39 & $\sim 143$ & MeV/c \\
final beam momentum after L1 & $\sim 350$ & MeV/c \\
\hline \hline
\end{tabular}
\end{center}
\end{table}
 \begin{figure}[hhhhhhhh!!!!!!!]
\centering
\includegraphics[width=0.50\textwidth]{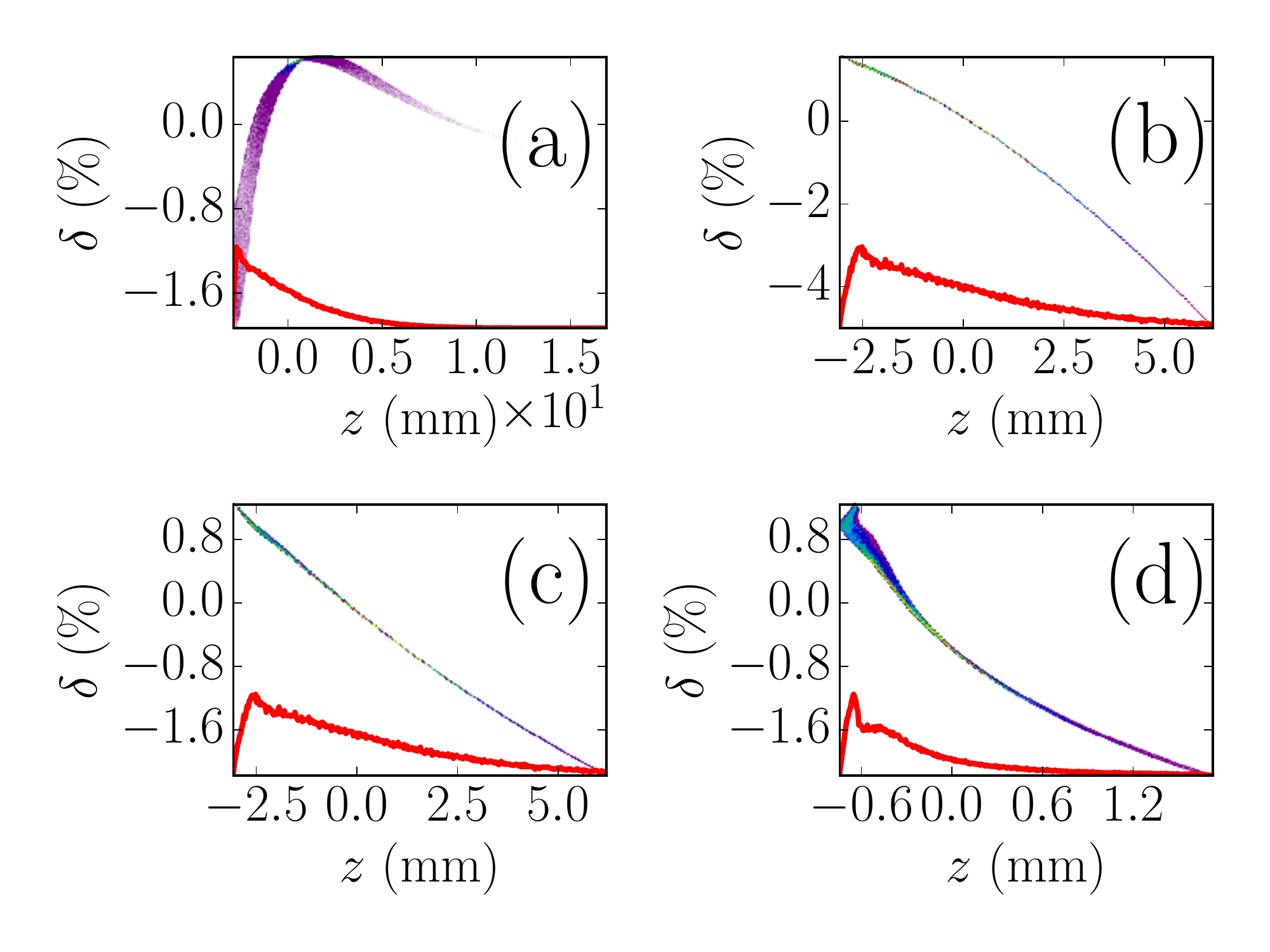}
\caption{Snapshots of the longitudinal phase spaces and associate current profiles (red traces) upstream of L0 (a) and downstream of L0 (b), L39 (c) and BC (d). Simulations up to L39 are 
carried with {\sc astra} whereas a one-dimensional longitudinal-dynamics model is used for BC2. The head of the bunch corresponds to $z>0$.
\label{fig:LPSsnap}}
\end{figure}

 Given the devised configuration, a  one-dimensional model of the longitudinal beam dynamics was employed to asses the viability of the 
 required compression and especially explore the possible impact of nonlinearities in the longitudinal phase space on the achieved current profile. We  
 considered the current could be longitudinally compressed using a conventional magnetic bunch compressor (BC) with longitudinal linear and second order 
dispersions $R_{56}$ and $T_{566}\equiv -\frac{3}{2}R_{56}$~\cite{emmaBC}. In our simulations the longitudinal dispersion was taken to $R_{56}= -20$~cm following similar designs~\cite{ttf2bc}. 
The phase of L0 and phase and amplitude of L39 were empirically optimized and 
the resulting longitudinal phase space $(z_0,\delta_0)$  was tracked through the BC via the transformation $z_0 \rightarrow z=z_0+R_{56}\delta_0+ T_{566}\delta_0^2$. 
An optimum set of phases and amplitudes was found and listed in Tab.~\ref{tab:dwfaFEL} and the sequence of the longitudinal phase spaces along the injector 
appear in Fig.~\ref{fig:LPSsnap}.  The  final wakefield excited in the structure with parameters listed in Table~\ref{tab:dlwperf} is displayed in Fig.~\ref{fig:finalwake} (b) $-$  the achieved field  and transformer ratio values are summarized in Table~\ref{tab:dlwperf}. We remark that the inclusion of a refined model of longitudinal dynamics leads to the apparition of features 
[e.g. a small current spike in the bunch tail; see Fig.~\ref{fig:finalwake} (b) or ~\ref{fig:LPSsnap} (d)] that were absent 
in the optimization process implementing a simple scaling of the longitudinal coordinates; see Fig.~\ref{fig:finalwake} (a). The origin of the small current spike can be traced back to the nonlinear correlation imposed by space charge in the early stages of the bunch-transport process (i.e. in the drift space upstream of L0); see Fig.~\ref{fig:LPSsnap} (a). Nevertheless the 
achieved peak field and transformer ratio as the bunch passes through the DLW are very close (within 10\%) to the ones obtained with the scaled distribution. 
These results indicate that our proposed injector concept appears to produce the required current profile. 
Further studies, including a transverse beam dynamics optimization and the inclusion of collective effects such as coherent synchrotron radiation and space charge downstream of L39 
and throughout the bunch compressor, will be needed to formulate a detailed design of the injector.  We nevertheless stress that the simple model presented above confirms a plausible longitudinal-beam-dynamics capable of preserving the formed current profiles after acceleration and compression. The final energies and peak currents are all within the parameters suggested in Ref.~\cite{zholents14}. 
 \begin{table}[hhh!]
\caption{\label{tab:dlwperf} Dielectric-line waveguide (DLW) parameters and resulting wakefield values using the current profile 
shown in Fig~\ref{fig:finalwake}. The ``ideal-" and ``realistic-compression" entries  respectively correspond to the cases when the 
final current profile is obtained via a simple longitudinal-axis scaling or via particle tracking. }
\begin{center}
\begin{tabular}{l c c  }
\hline \hline  parameter, symbol         &   value       & units  \\ \hline
DLW inner radius,   $r_i$    &    750 &  $\mu$m    \\
DLW outer radius,  $r_o$   &    795  &  $\mu$m      \\
DLW relative permittivity, $\epsilon_r$    &    5.7 &  --    \\
DLW fundamental mode, $f_1$ & 369.3   & GHz \\ 
\hline
{\bf ideal compression:} &&\\
Peak decelerating field, $|E_-|$     &    14.01 &  MV/m    \\
Peak accelerating field, $|E_+|$     &    75.55 &  MV/m    \\
transformer ratio, ${\cal R}$ & 5.39 \\
{\bf realistic compression:} &&\\
Peak decelerating field, $|E_-|$     &    12.84 &  MV/m    \\
Peak accelerating field, $|E_+|$     &    63.87 &  MV/m    \\
transformer ratio, ${\cal R}$ & 4.95 \\
\hline \hline
\end{tabular}
\end{center}
\end{table}

~~~\\
We finally note that the generated current profiles are capable of supporting electric fields and transformer-ratios in a DLW structure with performances that strike a balance 
between the two cases listed as  ``case 1" and ``case 2" in Table~1 of Ref.~\cite{zholents14}; see Tab.~\ref{tab:dlwperf}. A simple estimate indicates that our drive bunch 
would require a DWFA linac of $\sim 30$~m in order to accelerate an incoming 350-MeV witness bunch to a final energy of $\sim 2$~GeV. 

\section{Summary}
In conclusion, we have presented a set of smooth current profiles for beam-driven acceleration which display comparable performances with more complex 
discontinuous shapes discussed in previous work. We find that all proposed current profiles which lead to uniform decelerating fields ``live'' on the same performance curve
and that a given profile can be scaled to a particular accelerating field or transformer ratio. 
We also presented a simple laser-shaping technique combined with a photoinjector to generate our proposed quadratic current profile. 
We finally illustrated the possible use of this technique to  form an electron bunch with a tailored current profile. 
The distribution obtained from these start-to-end simulations were shown  to result in a transformer ratio $\sim 5$  
and peak accelerating field of $E_+\sim 60$~MV/m in a dielectric-lined waveguide consistent with the proposal of Ref.~\cite{zholents14}.  The method offers greater 
simplicity over other proposed techniques, e.g., based on complex phase-space manipulations~\cite{piotPRLshaping,piotprstab11}. Finally, we point out that the proposed 
method could provide bunch shapes consistent with those required to mitigate energy-spread and transverse emittance dilutions due coherent-synchrotron-radiation in 
magnetic bunch compressors~\cite{chad}.  \\


%
\section{Acknowledgments}

We would like to acknowledge members of the  ANL-LANL-NIU working group on DWFA-based short wavelength FEL led by J. G. Power and A. Zholents for useful discussions  
that motivated the study presented in this paper. P.P. thanks R. Legg (Jefferson Lab) and J. Bisognano (U. Wisconsin) for providing the 200-MHz quarter-wave field map used 
in Section~\ref{sec:linac}. This work was supported by the U.S. Department of Energy contract No. DE-SC0011831 to Northern Illinois University, and the Defense Threat Reduction Agency, Basic Research Award \# HDTRA1-10-1-0051, to Northern Illinois University. P.P. work is also supported by the U.S. Department of Energy under contract  DE-AC02-07CH11359 with the Fermi Research Alliance, LLC, and F.L. was partially supported by a dissertation-completion award granted by the Graduate School of Northern Illinois University. \\

\appendix
\section{Maximum of $A\cos(kz)+B\sin(kz)$ \label{app:A}}
The accelerating field behind the bunch often assumes the functional form 
\begin{eqnarray}\label{eq:cossin}
F(z)&=& A\cos (kz) + B \sin(kz). 
\end{eqnarray}
The procedure to evaluate the transformer ratio entails to determining the maximum value of $F(z)$. Such a value if 
found by solving for
\begin{eqnarray}
\frac{dF(z)}{dz}&=& k[-A\sin (kz)+B\cos (kz)]=0, 
\end{eqnarray}
with solution $z_s$ given by 
\begin{eqnarray}
\tan (kz_s)=\frac{B}{A}\equiv T. 
\end{eqnarray}
Squaring the previous equation, it is straightforward to show that 
\begin{eqnarray}
\sin^2 (kz_s)=\frac{ T^2}{1+T^2},\mbox{~and } \cos^2 (kz_s)=\frac{1}{1+T^2}. 
\end{eqnarray}

Expressing the value of $F(z_s)$ using the previous equation in~\ref{eq:cossin} leads to the 
maximum value of $F(z)$
\begin{eqnarray}\label{eq:final}
\hat{F}\equiv F(z_s)&=& A\sqrt{1+T^2}. 
\end{eqnarray}
The latter equation is used at several instances throughout  Section~\ref{sec:theory}. \\

\section{Analytic descriptions of the linear-ramp and double-triangle distributions \label{app:B}} 
In this Appendix we summarize and rewrite in notations consistent with our Section~\ref{sec:theory} the equations  
describing the linear ramp~\cite{bane} and double-triangle~\cite{jiang} current profiles. These 
equations are the ones used in Section~\ref{sec:theory-comp}.

The ``doorstep'' current profile considered in Ref.~\cite{bane} is written as 

\begin{equation}\label{eq:doorstep}
S(z)= 
\begin{cases} a  & \text{if $0 \le z < \xi$,} 
\\
a(\frac{2\pi(z-\xi)}{\lambda}+1) & \text{if $\xi\le z \le Z $,}
\\ 0 &\text{elsewhere.}
\end{cases}
\end{equation}

The ``double-triangle'' suggested in Ref.~\cite{jiang} is given in our notations as 

\begin{equation}\label{eq:doorstep}
S(z)= 
\begin{cases} akz  & \text{if $0 \le z < \xi$,} 
\\
a(kz-1) & \text{if $\xi\le z \le Z $,}
\\ 0 &\text{elsewhere.}
\end{cases}
\end{equation}

For both cases, $\xi=\lambda/4$ leads to the flat decelerating fields over the tail of distribution and leads to the $E_-^m$ and
${\cal R}$ tabulated and illustrated in Tab.~\ref{tab:Rcompare} and Fig.~\ref{fig:EvRanalytic} respectively.

%
%


\begin{thebibliography}{99}   
%
\bibitem{pisin} P. Chen, J.M. Dawson, Robert W. Huff, T. Katsouleas, Phys. Rev. Lett. {\bf 54}, 693 (1985). 
\bibitem{petra} G. A. Voss, and T. Weiland, ``Particle acceleration by wakefields", report DESY-M-82-10 available from DESY Hamburg (1982).
\bibitem{gai} W. Gai, P. Schoessow , B. Cole, R. Konecny, J. Norem, J. Rosenzweig, and J. Simpson, Phys. Rev. Lett. {\bf 61},  2756 (1988). 
\bibitem{wake} R. D. Ruth, A. Chao, P. L. Morton, P. B. Wilson, Part. Accel. {\bf 17}, 171 (1985). 
\bibitem{tsakanov} V. V. Tsakanov, Nucl. Instrum. Meth. Phys. Res. Sec. A, {\bf 432}, 202- (1999).
\bibitem{jing} C. Jing, A. Kanareykin, J. G. Power, M. Conde, Z. Yusof, P. Schoessow, and W. Gai, Phys. Rev. Lett. {\bf 98}, 144801 (2007). 
\bibitem{powersTR} J.G. Power, W. Gai, and P. Schoessow, Phys. Rev. E, {\bf 60}, 6061 (1999). 
\bibitem{bane} K. L. F. Bane, P. Chen, P. B. Wilson, ``On collinear wakefield acceleration," SLAC-PUB-3662 (1985).
\bibitem{jiang} B. Jiang, C. Jing, P. Schoessow, J. Power, and W. Gai, Phys. Rev. ST Accel. Beams {\bf 15}, 011301 (2012). 
\bibitem{muggli} P. Muggli, V. Yakimenko, M. Babzien, E. Kallos, and K. P. Kusche, Phy. Rev. Lett. {\bf 101}, 054801 (2008).
\bibitem{piotprstab11} P. Piot, Y.-E Sun, J. G. Power, and M. Rihaoui, Phys. Rev. ST Accel. Beams {\bf 14}, 022801 (2011). 
\bibitem{yinesun} Y.-E Sun, P. Piot, A. Johnson, A. H. Lumpkin, T. J. Maxwell, J. Ruan, and R. Thurman-Keup,  {Phys. Rev. Lett.} {\bf  105}, 234801 (2010).
\bibitem{powerIPAC14} G. Ha, M.E. Conde, W. Gai, C.-J. Jing, K.-J. Kim, J.G. Power, A. Zholents, M.-H. Cho, W. Namkung, C.-J. Jing, in Proceedings of the 2014 International Particle Accelerator Conference (IPAC14), Dresden Germany, 1506 (2014). 
\bibitem{zholents14} A. Zholents, W. Gai, R. Limberg, J. G. Power, Y.-E Sun, C. Jing, A. Kanareykin, C. Li, C. X. Tang, D. Yu Shchegolkov, E. I. Simakov, ``A collinear wakefield accelerator for a high-repetition-rate multi-beamline soft X-ray FEL facility," in Proceedings of the 2014 Free-Electron Laser conference (FEL14), 993 (2014).
\bibitem{wilson} A. Chao, {\em Physics of Collective Instabilities in High-Energy Accelerators}, Wiley Series in Beams \& Accelerator Technologies, 
John Wiley and Sons (1993).
\bibitem{flemeryAAC14} F. Lemery and P. Piot, ``Alternative Shapes and Shaping Techniques for Enhanced Transformer Ratios in Beam Driven Techniques, " in Proceedings of the 16th Advanced Accelerator Concepts Workshop (AAC 2014), San Jose, CA, July 13-18, 2014 (in press); also Fermilab preprint FERMILAB-CONF-14-365-AD (2014). 
\bibitem{andonian} G. Andonian, ``Title Holder, " in Proceedings of the 16th Advanced Accelerator Concepts Workshop (AAC 2014), San Jose, CA, July 13-18, 2014 (in press). 
\bibitem{lemeryshape} F. Lemery, D. Mihalcea, and P. Piot, in Proceedings of IPAC2012, New Orleans, Louisiana, USA, 3012 (2012). 
\bibitem{slacgun} D. T. Palmer, R. H. Miller, H. Winick, X.J. Wang,  K. Batchelor, M. Woodle, and I. Ben-Zvi, ``Microwave measurements of the BNL/SLAC/UCLA 1.6-cell photocathode RF gun", in Proceedings of the 1995 Particle Accelerator Conference, PAC'95 (Dallas, TX, 1995), 982 (1996).
\bibitem{astra}  K. Fl\"ottmann, {\em {\sc astra}: A space charge algorithm, User's Manual}, available from the world wide web at  http://www.desy.de/$\sim$mpyflo/AstraDokumentation (unpublished).
\bibitem{flIPAC14} F. Lemery and P. Piot, n Proceedings of the 2014 International Particle Accelerator Conference (IPAC14), Dresden Germany, 1454 (2014).
%
\bibitem{luiten} O. J. Luiten, S. B. van der Geer, M. J. de Loos, F. B. Kiewiet, and M. J. van der Wiel
Phys. Rev. Lett. {\bf 93}, 094802 (2004). 
%
\bibitem{wiener} A. M. Wiener, Rev. Sci. Instrum. {\bf 71}, 1929 (2000).
\bibitem{gibbs} J. W. Gibbs, Nature {\bf 59} (1539), 606  (1899). 
\bibitem{piotellipsoidal} P. Piot, Y.-E Sun, T. J. Maxwell, J. Ruan, E. Secchi, J. C. T. Thangaraj,  Phys. Rev. ST Accel. Beams {\bf 16}, 010102 (2013).
\bibitem{ferrini} G. Ferrini, P. Michelato, and F. Parmigiani, Solid State Commun. {\bf 106}, 21 (1998).
%
\bibitem{penco1} G. Penco, M. Trov\'o~ and S. M. Lidia, in Proceedings of FEL 2006, BESSY, Berlin, Germany, 621 (2006). 
\bibitem{penco2} G. Penco, M. Danailov, A. Demidovich, E. Allaria, G. De Ninno, S. Di Mitri, W. M. Fawley, E. Ferrari, L. Giannessi, and M. Trov\'o, 
Phys. Rev. Lett. {\bf 112}, 044801 (2014). 
%
%
\bibitem{aunes} B. Aunes, et al., Phys. Rev. ST Accel. Beams {\bf 3}, 092001 (2000).
\bibitem{bobgun} R. Legg, W. Graves, T. Grimm, and P. Piot, in Proceedings of the 2008 European Particle Accelerator Conference (EPAC08), Genoa, Italy, 469 (2008). 
\bibitem{JoeResults} J. Bisognano, M. Bissen, R. Bosch, M. Efremov, D. Eisert, M. Fisher, M. Green, K. Jacobs, R. Keil, K. Kleman, G. Rogers, 
M. Severson, D. D. Yavuz, R. Legg, R. Bachimanchi, C. Hovater, T. Plawski, T. Powers, in Proceedings of the 2013 North-American  Particle Accelerator Conference  
(NAPAC'13), Pasadena, USA, 622 (2013). 
\bibitem{WiFEL} J. Bisognano, R. A. Bosch, D. Eisert, M. V. Fisher, M. A. Green, K. Jacobs, K. J. Kleman, J. Kulpin, G. C. R. Edit, in Proceedings of 2011 
Particle Accelerator Conference (PAC11), New York, NY, USA, 2444 (2011). 
\bibitem{piotTESLANote} K. Fl\"ottmann, T. Limberg, and P. Piot, " Generation of Ultrashort Electron Bunches by Cancellation of Nonlinear Distortions in the 
Longitudinal Phase Space," DESY report TESLA FEL 2001-06, available from DESY, Hamburg Germany (2001). 
\bibitem{39GHz} N. Solyak, I. Gonin, H. Edwards, M. Foley, T. Khabiboulline, D. Mitchell, J. Reid, L. Simmons, in Proceedings of the 2003 Particle Accelerator 
Conference (PAC03), Portland, OR, USA, 1213 (2003).
%
\bibitem{geneticoptimizer} M. Borland, and H. Shang, {\sc geneticOptimizer}, private communication (2005). 
\bibitem{emmaBC} P. Emma, ``Bunch Compressor Options for the New TESLA Parameters", Internal unpublished report DAPNIA/SEA-98-54 available from Service des Acc\'el\'e rateurs, CEA Saclay, France (1998)
\bibitem{ttf2bc} T. Limberg, Ph. Piot and F. Stulle, in Proceedings of the 2002 European Particle Accelerator Conference (EPAC2002), Paris France, 1544 (2002).
\bibitem{piotPRLshaping} P. Piot, C. Behrens, C. Gerth, M. Dohlus, F. Lemery, D. Mihalcea, P. Stoltz, M. Vogt, Phys. Rev. Lett. {\bf 108}, 034801 (2012).
%
\bibitem{rosinggai} M.~Rosing, and W.~Gai, Phys. Rev. D {\bf {42}}, 1829 (1990).
 \bibitem{chad} C. Mitchell, J. Qiang, and P. Emma, Phys. Rev. ST Accel. Beams {\bf 16}, 060703 (2013). 
%
\end{thebibliography}
\end{document}